\documentclass[preprint,12pt]{elsarticle}

\journal{Physics Letters A}

\begin{document}

\begin{frontmatter}

\title{Laser Singular Theta-Pinch.}

\author{A.Yu.Okulov}

\address{General Physics Institute of 
Russian Academy of Sciences 
Vavilova str. 38, 119991, Moscow, Russia}

\begin{abstract}
The interaction of the two counter-propagating 
ultrashort laser pulses 
with a singular wavefronts in the thin slice of the 
underdense plasma is considered. 
It is shown that ion-acoustic wave is 
excited via Brillouin three-wave resonance by corkscrew interference pattern 
of a paraxial singular laser beams. 
The orbital angular momentum carried by light is transferred to 
plasma ion-acoustic vortex. The rotation of the density perturbations 
of electron fluid is the cause of helical current which produce the 
kilogauss axial 
quasi-static magnetic field. 
The exact analytical configurations are presented for an 
ion-acoustic current field and magnetic induction. The range 
of experimentally accessible parameters is evaluated.
\end{abstract}


\end{frontmatter}

\section{Introduction}
The Stimulated Brillouin Scattering (SBS) in a plasma \cite{Alexandrov:1984,Kruer:1990}
is a subject of the considerable 
interest 
in a several recent decades \cite{Rosenbluth:1974}. 
The motivation of these research efforts is that 
a substantial power could be reflected from an underdense plasma via SBS thus 
limiting the absorption of laser radiation in an inertial confinement fusion 
(ICF) targets \cite{Lushnikov:2004}.
The possibilities of the ion-acoustic plasma wave resonant seeding by a controllable 
interference patterns of crossed laser beams \cite{Baldis:1996} were 
studied in order to manipulate the SBS and Raman scattering \cite{Kirkwood:1997}.
On the other hand the SBS itself and in combination 
with a random phase plates (RPP) is considered as a possible tool 
for laser beam 
smoothing and quenching of spatial instabilities in laser plasma 
\cite{Tikhonchuk:1997,Tikhonchuk:2001,Loiseau:2006}. The chaotic 
intensity and phase variations in a speckle pattern induced 
by RPP are also in use 
in wavefront reversal SBS mirrors in order to improve 
phase-conjugated (PC) replica fidelity \cite{Basov:1980,Zeldovich:1985}. 
As is shown recently the dark lines of a speckle (optical vortices) 
which are collocated with the optical phase singularities 
\cite{Berry:1974} inside SBS mirror 
volume are surrounded by the corkscrew interference 
patterns \cite{Okulov:2008,Okulov:2009}. 
These optical helical patterns rotate synchronously 
with an acoustic angular frequency $\Omega_{ia}$ and their rotation 
is accompanied by a circular motion of the SBS medium inside PC-mirror \cite{Okulov:2008J}. 
Such a circular motion of the 
medium carries an orbital angular momentum (OAM) 
\cite{Thomas:2008} 
extracted from the exciting radiation \cite{Okulov:2008J}. In a similar 
way the transfer of OAM to an ensemble of $10^4$ ultracold cesium atoms 
\cite{Tabosa:1999} via rotating interference pattern 
produced by a pair of the 
singular Laquerre-Gaussian(LG) laser 
beams had been observed in a nondegenerate four-wave mixing PC geometry. 
Quite recently the else example of an optical 
corkscrew patterns had been demonstrated with photorefractive 
Ba:TiO$_3$ PC mirror \cite{Woerdemann:2009}.

The goal of this paper is in analysis of a new mechanism of an ion-acoustic wave (IAW)
management by means of seeding the corkscrew interference pattern rotating around 
the isolated optical phase singularity. The qualitative picture is 
as follows. The optical helical pattern rotates due to a frequency 
detuning $\Omega_p - \Omega_s $ between the pump and Stockes 
LG beams (fig.\ref{fig.1}) 
which collide inside freely expanding preformed underdense 
plasma jet (PJ, fig.\ref{fig.2}) created by evaporation of 
the solid-state target by nanosecond laser pulse. The 
parameters are expected to be 
close to experimental conditions 
summarized in \cite{Tikhonchuk:1997}. The equilibrium atomic 
density $n_{i,e}$ is in the range $(0.2-0.07){\:}n_{c}$ where 
$n_c = {1.011 \cdot 10^{21} cm^{-3}}$ is the critical density for laser 
wavelength $\lambda=1.05 \mu m$ \cite{Tikhonchuk:1997}. 
We suppose the detuning is adjusted in resonance 
with IAW having frequency $\Omega_{ia} \approx 2 \pi \cdot 10^{12}$
and fast damping rate 
$\gamma_{ia} \approx (0.5-0.1)\Omega_{ia}$ 
\cite{Kirkwood:1997,Tikhonchuk:1997}. 
In this SBS strong damping limit when duration of laser pulse is 
about $\tau \approx 10^{-10-11} sec$ the helical density perturbations 
of the electron ($\delta n_{e}$) and ion ($\delta n_{i}$) liquids 
of the order 
of $10^{-2} n_{i,e}$ are collocated with the light intensity maxima.
The rotation of the scalar density fields $\delta n_{i,e}$ produces 
a $nonrelativistic$ helical plasma flow with angular 
frequency $\Omega_{ia}$. This helical current 
in the neutral plasma reminiscent to the geometry of a $\Theta-pinch$ 
\cite{Alexandrov:1984,Kruer:1990} is 
expected to produce the axial quasi-static 
magnetic field. The mechanism discussed here is different from the 
process of the photon's 
spin deposit from the circularly polarized intense light pulse to 
underdense plasma rotation 
via inverse Faraday effect described in \cite{Haines:2001,Bychenkov:1994} 
and already identified experimentally 
in \cite{Najmudin:2001} by measurement 
the megagauss (MG) $axial$ magnetic induction ${\bf B}_z$.
In our case the $linearly$ $polarized$, i.e. $zero$ $spin$ LG photons
\cite{Barnett:2002} transfer their OAM to an 
ion-acoustic liquid via three-wave Brillouin resonance. The other 
difference is in the much lower (kilogauss range) magnitude 
of axial magnetic induction ${\bf B}_{z}$  resulting in the absence 
of the so-called IAW curtailment \cite{Haines:2001}. 
The resulting helical flow 
of electron liquid (fig.\ref{fig.1}) generates 
both $axial$ and $azimuthal$ magnetic 
inductions ${\bf B}_z$ and ${\bf B}_{\theta}$. 
We will show that 
interference pattern (fig.\ref{fig.1}) rotates as a 
"solid body" and 
the vector of axial speed ${\bf v}_z$ (helix $pitch$) 
is ${\lambda}/4 {\pi}R$ times 
shorter than the vector of azimuthal rotation speed ${\bf v}_{\theta}$. 
Thus the same ratio holds for the current density vectors ${\bf j}_{z}$ 
and ${\bf j}_{\theta}$. This leads to the corresponding 
increase of the axial component of magnetic field 
${\bf B}_{z}$ compared to azimuthal one ${\bf B}_{\theta}$. 

The most interesting issue is the interaction of electron 
and ion liquids in 
the process of SBS excitation by a short laser pulse having 
picosecond duration ($\tau \approx 10^{-10-11}sec$). The 
standard assumption of the plasma quasi-neutrality 
on the scales larger than Debye 
length ($r_D= \sqrt{k_B T_e \epsilon_0/e^2 n_e}\sim 10^{-6-7}cm$) requires 
the instantaneous compensation of the 
each electron density perturbation by an appropriate motion 
of the ion liquid. 
Thus within a framework of the conventional 
IAW model the macroscopic electron current must be exactly 
compensated by ionic current and net current ought to be equal to zero 
resulting in zero magnetic induction.  
This simple assumption is a basement of the three-wave SBS model (3 scalar 
"parabolic" PDE) 
elaborated previously for the subnanosecond plasma flow 
with $KeV$ temperatures \cite{Loiseau:2006}. 
Let us look more carefully to this assumption from the point of 
view of the optical pulse durations in the range 
mentioned above ($10 - 500$ picoseconds): 
both the experiments followed by analytical 
estimates\cite{Tikhonchuk:1997} and numerical 
simulations \cite{Loiseau:2006} 
supposed no magnetic fields. But the slight change of duration $\tau$ down 
to the $1$ picosecond revealed the substantial magnetic 
fields \cite{Borghesi:1998,Borghesi:1998P}. Despite 
the fact of a five order 
increase of the optical fluxes (up to $10^{19} W/cm^2$) the sudden 
appearence of a MG magnetic field for a $1 psec$ flows requires 
proper interpretation. The possible interpretation is that a 
Debye screening do not cancel the net current completely 
due to partial imbalance in between the ion ($n_i$) and 
electron ($n_e$) liquids. As a result IAW is transformed
into magnetosonic wave \cite{Alexandrov:1984,Haines:2001}. 
In order to simplify consideration we will use 
the steady-state vortex solutions of SBS IAW with 
phenomenological parameter $\eta \sim 0.01 - 0.1$ which 
is the ratio of imbalanced electron current compared to 
ionic current. The advantage of this simple assumption is 
in possibility of getting the exact field of velocities and 
density profile without solving 
kinetic equation.

Noteworthy the other mechanisms of magnetic fields 
generation. For example the $axial$ currents co-directed with 
picosecond laser pulse were identified 
experimentally by Faraday rotation polarimetry as a source of 
the toroidal magnetic field of the MG range 
 \cite{Borghesi:1998,Borghesi:1998P}. 
Laser $Z-pinch$ formed by return current of the hot 
electrons ejected from a thin 
copper wire by an ultrashort laser pulse 
was studied in \cite{Beg:2004}. 
The Weibel instability mechanism is responsible for generation of the 
quasistatic magnetic fields for the anisotropic electron's density in both 
nonrelativitic helical plasma flows \cite{Silin:1990} 
and intense femtosecond laser pulse in an underdense 
plasma \cite{Krainov:2003}. The singularities in density and magnetic 
field growing in time as $(t-t_0)^{-1/2}$  
were considered earlier in \cite{Silin:1990}. 
The gigagauss magnetic induction $\bf B$ mechanisms were 
discussed in \cite{Sudan:1993,Krainov:2008,Beg:2004}. 

\section{Helical interference patterns of the 
Laguerre-Gaussian beams and angular momentum density}
SBS is a decay of the quasimonochromatic 
electromagnetic wave ${\bf E}_{p}$ into the 
Stockes wave ${\bf E}_{s}$ and the 
ion-acoustic wave $Q$ via Bragg scattering. 
The mechanism of this three-wave parametric instability 
\cite{Tikhonchuk:1997} is the electrostrictive excitation of a 
sound wave by the moving interference pattern of two 
waves (${\bf E}_{p}$  and ${\bf E}_{s}$) 
detuned by the sound frequency $\Omega_{ia} = 2\Omega_p {\:}n_r c_{ia}/c$ 
\cite{Alexandrov:1984,Kruer:1990,Zeldovich:1985}, where $c_{ia}$ is 
the velocity of the ion-acoustic wave, $n_r$ is refractive index. 
The underdense 
 plasma frequency $\omega_p = \sqrt {n_c {\:}e^2/\epsilon_0 m_e}$ 
is assumed to be 
smaller than laser frequency $\Omega_p \approx \omega_p / \sqrt{0.2}$. 

\begin{figure}
\center{\includegraphics[width=0.99\linewidth] {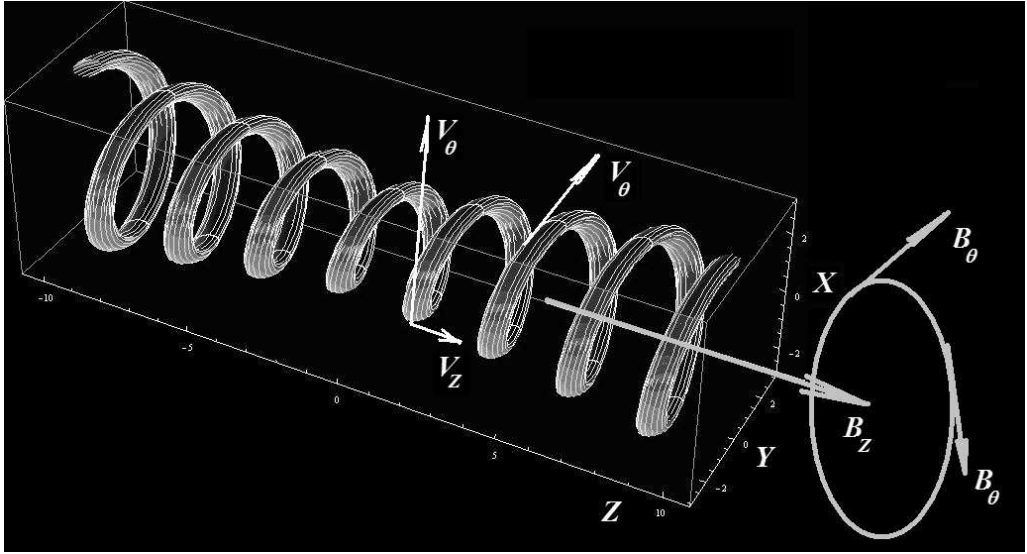}}
\caption{The isosurfaces of the corkscrew interference pattern 
\cite{Tsytovich:2007,Okulov:2008} of a two 
detuned counter-propagating 
phase-conjugated $LG^{1}_0$ $vortex$ laser beams inside plasma jet. 
(Only one helix of the two is shown.) The rotation of 
corkscrew pattern causes the motion of a density perturbation $\delta n_e$ with 
longitudinal speed ${\bf v}_z={\bf\vec z} c_{ia}$ and azimuthal speed 
$ {\bf v}_{\theta}=[{\Omega_{ia} {\bf\vec z}} \times {\vec R}]$. 
The solenoidal current produces dipole-like 
quasi-static magnetic field $\bf B$. The axial field component ${\bf B}_z$ is produced 
by an azimuthal current density ${\bf j}_{\theta}= 2 e {\bf v}_{\theta} \delta n_e$, 
while the axial current component ${\bf j}_{z}= 2 e {\bf v}_{z} \delta n_e$ generates an 
azimuthal field ${\bf B}_{\theta}$.}
\label{fig.1}
\end{figure}
In quantum terms the decay of the each pump photon to a 
Stockes photon and an 
ion-acoustic phonon follows to energy conservation condition 
$\hbar {\:} \Omega_{p} - \hbar {\:}\Omega_s = \hbar {\:} \Omega_{ia}$ 
and to momentum conservation 
$\hbar \vec k_{p} - \hbar \vec k_s = \hbar \vec k_{ia} \approx 2 \vec k_p$ as well. 
The angular momentum conservation \cite{Okulov:2008} 
also takes place resulting 
in OAM deposit in rotating electron liquid 
(rather than photon's spin deposit \cite{Haines:2001}).  
In classical picture the detuning between optical 
waves $\Omega_{ia}$ 
is a Doppler shift arising due to reflection 
from sound grating moving with the speed $c_{ia}$ in 
a medium (plasma) with refractive index 
$n_r$. 
For the backward SBS
the period of sound grating $p=\lambda_{ia}$
is approximately equal to a half of exciting wavelength $\lambda_{ia} \approx \lambda_p/2$ 
(Bragg reflection condition). 
Noteworthy the Bragg and Doppler 
conditions are valid both near the bright spots and in the vicinities 
of the dark lines as well. The fundamental difference of the dark 
(vortex) lines from the bright spots of the speckle stems from the different 
structure of their optical wavefronts. In a bright spot the wavefront 
is parabolic while near a vortex line the wavefront is 
helicoidal \cite{Berry:1974,Okulov:2008}.
In spite of the statistical nature of a speckle the complex amplitudes 
of the electric fields ${\bf E}_{p,s}$ for the 
counter-propagating waves 
could be written analytically inside a given bright or dark spot of a speckle.
For a homogeneous polarization state the electric field has the following form 
in a cylindrical coordinate system $(z,r,\theta,t)$ nested in a given spot:
\begin{eqnarray}
\label{electric_field}
\ {{{\bf E}}_{(p,s)}(z,r,\theta,t)}= 
{{{\bf E}}^{o}_{p,s}}{\:} f_{p,s}(z,r){\:} {\:}{r^{|\ell|}} {\:} 
&& \nonumber \\
\times {\:} exp  {\:} [ {\:} 
 i{\Omega_{(p,s)}}t \mp ik_{(p,s)}z +i{\chi}_{p,s}(z,r)\pm i{\ell}\theta {\:}], 
{\:}{\:}{\:}{\:} 
\end{eqnarray}
where $f_{p,s}(z,r)$ is a smooth amplitude function elongated in $z$-direction 
(e.g. Gaussian one), 
${\chi}_{p,s}(z,r)$ is a smooth phase profile (e.g. parabolic one),
$\ell$ is an azimuthal quantum number (topological charge).
Define the mean square root transverse scale of a 
speckle entity as ${<D>}$. Then mean longitudinal length of a variational 
speckle functions $f_{p,s}(z,r)$ and ${\chi}_{p,s}(z,r)$ 
is the Rayleigh range (Fresnel length) $L_R \approx {<D>}^2/{\lambda}$ 
\cite{Zeldovich:1985}. 
Both $f_{p,s}(z,r)$ and ${\chi}_{p,s}(z,r)$ are 
assumed to be parabolic with respect to $r$ near $z$-axis. For ${\ell}=0$ 
the equation (\ref{electric_field}) describes the 
phase-conjugation of a bright spot, e.g. zeroth-order Gaussian beam.
For ${\ell} \geq 1$ and 
when $f_{p}(z,r)=f_{s}(z,r)$ and ${\chi}_{p}(z,r)={\chi}_{s}(z,r)$ the 
eq.(\ref{electric_field}) describes the perfect phase-conjugation 
of an optical vortex with ultimate fidelity i.e. when the 
correlation integral of the pump ${\bf E}_{p}$ and Stockes ${\bf E}_{s}$ 
fields is equal to unity \cite{Okulov:2008J}. For the $\ell$-order 
$LG^{\ell}_0$ beam (LG) the $f_{p,s}(z,r)$ has Gaussian form \cite{Okulov:2008}. 
Thus two phase-conjugated LG compose in their common waist 
the corkscrew interference pattern 
(fig.\ref{fig.1}) \cite{Okulov:2008,Okulov:2008J}. Qualitatively 
the same corkscrew pattern in PC-mirror appears in the vicinity 
of the each optical vortex line of an optical speckle pattern:
\begin{eqnarray}
\label{inter_patt1}
|{{\bf E}_p(z,r,\theta,t)}+{{\bf E}_s(z,r,\theta,t)}|^2 = 
|{\bf E}^{o}_{p}|^2[1+{R_{pc}}+2{\sqrt {R_{pc}}}
&& \nonumber \\
{\:}{\:}cos[  {\:} (\Omega_p-\Omega_s) t - 
(k_p+k_s) z+2{\:}  {\ell}{\:}\theta {\:}]]{\:} 
\times{\:}r^{{\:}2{|\ell|}} {\:} {\:} f_{p,s}^{{\:}2}(z,r),
\end{eqnarray}
where $R_{pc}=|{\bf E}^{o}_{s}|^2/{|\bf E}^{o}_{p}|^2$ is the 
PC-reflectivity. The maximum of the interference pattern acts as an Archimedian screw 
transferring OAM from light to the matter \cite{Okulov:2008J}.
This happens because the orbital part of the 
electromagnetic angular momentum density $M_z$ of LG 
is collocated with a maximum of the light intensity \cite{Allen:1992}:
\begin{equation}
\label{oam_density}
\ M_z(z,r,\theta ,t) \approx {\frac {\ell}{\Omega_p}}|{\bf E}_{p,s}|^2+
{\frac {\sigma_z r }{2 {\:} \Omega_p }} {\:}
{\frac {\partial {|{\bf E}_{p,s}|^2} } {\partial r }},
\end{equation}
where $\sigma_z = 0, \pm 1$ corresponds to linear, right and left polarizations 
of LG respectively. Hence there is a principal 
difference in between a vortex line and a bright spot in the optical 
speckle. The bright spot only pushes the medium by 
$rolls$ of interference pattern \cite{Okulov:2008} along 
propagation direction ($z-axis$) of a pump wave ${\bf E}_{p}$, while the 
vortex line located in a dark spot produces the additional rotational effect upon 
liquid and imprints the angular momentum therein \cite{Okulov:2008J}. 

\section{Optical collider setup for ultrashort laser 
pulses with a singular wavefronts}
\begin{figure}
\center{\includegraphics[width=0.99\linewidth] {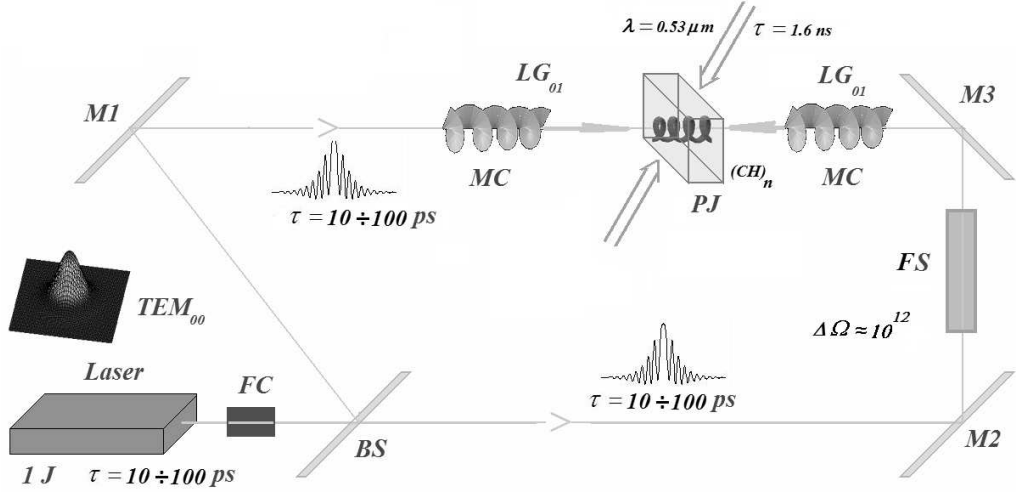}}
\caption{Conceptual optical loop setup for the colliding pulse SBS excitation in a preformed 
underdense $plasma$ $jet$ $\bf PJ$ \cite{Tikhonchuk:1997} produced 
by evaporation of carbon containing plastic film. The  
beam splitter $\bf BS$ divides $TEM_{00}$ gaussian laser beam into the two 
pulses. After the  passage through the 
$frequency$ $shifter$ $\bf FS$ and $mode$ $converters$ $\bf MC$ \cite{Allen:1992} 
the colliding pulses 
$LG^{1}_0$ with a singular wavefronts produce the rotating corkscrew 
pattern in an expanding $\bf PJ$. Under the proper matching of 
the rotation frequency $\Omega_p - \Omega_s$ and 
ion-acoustic frequency $\Omega_{ia}$ and under proper management of the 
path difference and wavefront alignment the rotating helical 
ion-acoustic wave is expected to be excited. $\bf M1,M2,M3$ are the 
mirrors, $\bf FC$ is a Faraday isolator.} 
\label{fig.2}
\end{figure}
The easiest way to produce the optical corkscrew by means of the 
interference of a pair of the two phase-conjugated 
optical vortices is to use a $loop$ optical setup (fig.\ref{fig.2}) 
reminiscent to optical collider schemes \cite{Bulanov:2006}. 
The Eq.(\ref{electric_field}) shows that a phase-conjugated optical 
vortex ${\bf E}_{s} \approx exp(+ik_{s}z - i{\ell} \theta {\:})$ has 
the same ratio of the signs of the $k_s$ and $\ell \theta$ in a self-similar variable 
as those of a pump wave ${\bf E}_{p} \approx exp(-ik_{p}z + i{\ell} \theta {\:})$. This 
means that a helical wavefront of the PC-vortex ${\bf E}_{s}$ has 
the same value and sign of the topological charge $\ell$, 
as an incident wave ${\bf E}_{p}$. Hence the PC-vortex is 
absolutely identical to an incident wave 
except for the exactly opposite direction of propagation. The restriction 
on the turnover of the direction of propagation is a number of reflections from 
conventional, i.e. non-PC mirrors. 
The $odd$ number of reflections changes the topological charge $\ell$ to the 
opposite one \cite{Okulov:2008}. Hence the conceptual $loop$ setup 
ought to perform the $even$ number of reflections in order to 
keep the topological charges of colliding pulses identical to each 
other as shown at fig.\ref{fig.2}. 
The alternative is to use 
two mode converters ($\bf MC$) at a final straight 
path before target $\bf PJ$ (also at fig.\ref{fig.2}). 
The other restriction imposed on $coherence$ $length$ means that a paths 
of the counter-propagating vortices 
ought to be equalized with an accuracy better than $L_{coh}=c {\:}\tau$ for 
proper temporal overlapping. 

For the transform-limited light pulses having 
$\tau \approx 1-100 {\:}ps$ duration the coherence length is 
$L_{coh} \approx 3 \times 10^3 -3 \times 10^5 \mu m$, i.e.  
above $300{\:} \cdot \lambda_p$. In such a case 
the ultrashort pulses with a singular wavefronts 
divided by a beamsplitter in a proposed loop scheme  
(fig.\ref{fig.2}) will collide in a thin slice of 
a preformed underdense ($\Omega_p > \omega_p=\sqrt {n_e e^2/m_e \epsilon_0}$) 
plasma jet \cite{Tikhonchuk:1997} 
and produce corkscrew interference pattern therein.
Because the SBS sound (IAW) grating parameters must 
be in a resonance defined by 
Doppler and Bragg conditions, the frequency shift between colliding pulses 
should be equal to the ion-acoustic 
frequency $\Omega_{ia}$ with an accuracy better 
than ion-acoustic SBS linewidth $\delta \Omega_{ia}$. Such frequency 
shift could be produced by means of Raman scattering in compressed gases.

The acoustical frequency $\Omega_{ia}$ is of the order of $10^{9} Hz$ for 
SBS in a room-temperature gases and liquids because it is determined by a thermal 
velocity $V_T \approx \sqrt{k_B T/m_{i,e}}$ at an ambient temperature $T \approx 300K$  
(0.025 $eV$), where $k_B$ is a Boltzmann constant and $m_{i,e}$ is a mass 
of a particle. 
The Brillouin linewidth in transparent dielectric is 
defined by a damping rate of sound \cite{Zeldovich:1985}. 
The ion-acoustic plasma frequency shift $\Omega_{ia}$ 
has roughly the same dependence on 
electron and ion temperatures $T_e$ and $T_i$. 
In this case the much higher $T_{i,e} \approx 1 keV$ offers the $10^{12}Hz$ 
frequency shift due to a basically same 
square root dependence $c_{ia} \approx \sqrt{k_B T_e/m_i}$, or more 
precisely $c_{ia}=\sqrt{k_B(Z \cdot T_e+3T_i)/m_i}$, where $Z$ is the ion's 
charge \cite{Loiseau:2006}. 
The SBS linewidth for IAW is due to the Landau 
damping $ \gamma_{ia} \sim \nu_L=\Omega_{ia} \sqrt {\pi Z m_e /8 m_i}$  \cite{Landau:1982}.

\section{Configuration of an ion-acoustic plasma vortex seeded by a corkscrew interference pattern}

Below are the SBS equations of motion for the scalar slowly varying envelope 
optical fields, i.e. ${{\mathcal {E}}}_{p}$ moving in the 
positive $z$-direction and ${\mathcal {E}}_{s}$ moving oppositely:
\begin{equation}
\label{pumpwave}
\ {\frac {\partial {{{{\mathcal {E}}}_p}(z,r,\phi,t )}} {\partial z} }+
{\frac {n_r} {c} }{\frac {\partial {{{{\mathcal {E}}}_p}}} {\partial t} }+
{\frac {i}{2 k_p}} {\nabla}_{\bot}^{{\:}2} {{{\mathcal {E}}}_p} =
{\frac {i \Omega_p {\:}{n_{e}}} {4{\:} c{\:}{n_c} } } {\tilde Q} {\:}{{\mathcal {E}}}_s 
\end{equation}
\begin{equation}
\label{stockeswave}
\ {\frac {\partial {{{{\mathcal {E}}}_s}(z,r,\phi,t )}} {\partial z} }-
{\frac {n_r} {c} }{\frac {\partial {{{{\mathcal {E}}}_s}}} {\partial t} }-
{\frac {i}{2 k_p}} {\nabla}_{\bot}^{{\:}2} {{{\mathcal {E}}}_s} = -
{\frac {i \Omega_s {\:}{n_{e}}} {4{\:} c{\:}{n_c} } } {{\mathcal {E}}}_p {{\tilde Q}}^{\ast},
\end{equation}
and dimensionless slowly varying IAW density 
perturbation complex 
amplitude ${\tilde Q}$ 
\cite{Alexandrov:1984,Kruer:1990,Lushnikov:2004,Tikhonchuk:1997,Loiseau:2006}:
\begin{eqnarray}
\label{acouswave1}
\ {\frac {\partial {{\tilde Q}(z,r,\phi,t )}} {\partial z} }+
{\frac {1} {c_{ia}}} {\frac {\partial {\tilde Q}}  {{\:}\partial t} }
+{\frac {2 {\gamma_{ia}} {{\tilde Q}}} {c_{ia}} }
+{\frac {i}{2 (k_p+k_s)}} {\nabla}_{\bot}^{{\:}2} {\tilde Q}=
&& \nonumber \\
{i(k_p+k_s) } {{{\mathcal {E}}}_p}  
{{{\mathcal {E}}}_s }^{\ast}{\frac {\epsilon_0 }{2{\:} {n_c}{\:}k_B T_{e}}},
\end{eqnarray}
where  ${\nabla}_{\bot}=({\partial_x},{\partial_y})$, 
$n_c$ is the critical plasma density \cite{Lushnikov:2004,Loiseau:2006}. 
For a preformed plasma jet of the thickness $L_{jet} \approx 10^3 \mu m$ \cite{Tikhonchuk:1997}
probed by two colliding singular laser 
beams  of comparable amplitudes ${\mathcal {E}}_p \approx {\mathcal {E}}_s$
(fig.\ref{fig.2}) with beam waist diameter $D \approx 100 \mu m$
and wavelength $\lambda_p \approx 1.05 \mu m$ 
the Rayleigh range is $L_R=k_p D^2 \approx 10^4 \mu m \approx 10 L_{jet}$. 
Then $\nabla_{\bot}^{{\:}2}$ terms which are responsible for the tranverse 
effects could be omitted. In the linear regime of SBS and under the 
weak coupling conditions the 
fields ${\mathcal {E}}_p$ and ${\mathcal {E}}_s$ are close to those fixed 
at their free-space values at the opposite 
boundaries of the jet $\bf PJ$ and their z-dependence is obtained exactly 
throughout all inner $z$. This weak-coupling regime might be justified 
due to nonsaturated SBS reflectivity 
($R_{sbs} \approx 0.03$) obtained in \cite{Tikhonchuk:1997}.
Next assumption is the stationary SBS regime 
when IAW damping is assumed to be 
strong: 
 $\gamma_{ia} \approx (0.5-0.1)\Omega_{ia}> {\tau}^{-1} \approx 10^{10-11} Hz$ 
\cite{Kirkwood:1997,Tikhonchuk:1997}.
Thus exact expression for the complex amplitude of IAW 
${Q}(z,r,\theta,t) ={\tilde {Q}} \cdot exp[i (\Omega_{ia}t-k_{ia}z)])$ 
follows immediately for the pair of colliding $LG$ vortices \cite{Okulov:2008} 
with $|{{\mathcal {E}}}^{o}_{p}|=|{{\mathcal {E}}}^{o}_{s}|$:
\begin{eqnarray}
\label{density_profile}
\ {Q} =exp{\:} [i(\Omega_{ia} t -k_{ia} z)+i2{\ell}\theta {\:}]{\:}
{\:}{(r/D)}^{{\:}2{|\ell|}}{\:}
&& \nonumber \\
exp  {\:} [ {\:} - {\frac {2{\:} r^2}{D^2(1+z^2/(k_p^2 {\:} D^4))}}]
 {\frac { \Omega_{ia}}{4 \nu_L }}{\frac {{\epsilon_0 }}{{n_c}{\:}k_B T_{e}}}|{\bf E}^{o}_{p}|^2.
\end{eqnarray}
Consequently the density perturbation field $q(z,r,\theta,t)={\delta n}/ {n_{e}}=Re{(Q)}$ 
demonstrates a rotation typical to optical vortex, alike LG 
beam. The difference is in doubled topological charge $2\ell$ 
due to the angular momentum conservation 
\cite{Okulov:2008,Okulov:2008J}:
\begin{eqnarray}
\label{density_profile_arb}
\ {q} =cos{\:} [\Omega_{ia} t -k_{ia} z+2{\ell} \theta {\:}]{\:}
{\:}{(r/D)}^{{\:}2{|\ell|}}{\:}
&& \nonumber \\
exp  {\:} [ {\:} - {\frac {2{\:} r^2}{D^2(1+z^2/(k_p^2 {\:} D^4))}}]
 {\frac { \Omega_{ia}}{4 \nu_L }}{\frac {{\epsilon_0}}{{n_c}{\:}k_B T_{e}}}|{\bf E}^{o}_{p}|^2,
\end{eqnarray}
Thus we see that in the colliding pulse geometry the IAW density 
perturbation $q$ has a form of a double helix which rotates around $z-axis$ 
with the angular frequency $\Omega_{ia}$ (fig.\ref{fig.1}).
Then an analytical expression for the ion-acoustic current 
density ${{\bf j}_{ia}}$ could be extracted 
from the dynamical equations for the electron liquid \cite{Alexandrov:1984,Kruer:1990}: 
\begin{equation}
\label{contin_eq}
{\frac {\partial {q}}{  \partial {\:}t}} + \nabla \cdot {q} {\bf V} = 0 ;
{\frac {d {\bf V}}{ d {\:}t}}  = -
{\frac  e {m_e}}{\:}[{\bf E} + [{\bf V} \times {\bf B}]]
\end{equation}
From the first equation (\ref{contin_eq})(continuity equation) the approximate 
expression for the ion-acoustic current density vector 
field ${\bf  j}_{ia}({\vec r},t)$ may be obtained using following symmetry arguments.
The $\dot q$ plays the role of the density of effective "charge" density 
of the long "charged twisted wire", 
while $q {\bf V}$ is an effective "electric field" directed 
along the normal to this "charged twisted wire" 
(fig.\ref{fig.1}, ${\bf x,y,z}$  are unit vectors). 
Then the current density is obtained via Gauss theorem:
\begin{eqnarray}
\label{current field}
{{\bf j}_{ia}({\vec r},t)}={2\eta \ell}{Z{\cdot}e{\cdot}n_e{\:}{q({\vec r},t)} {\bf V}} \approx
{2\eta \ell}{Z{\cdot}e{\cdot}n_e{\:}{q}({\vec r},t){\cdot}}
&& \nonumber \\
{[\vec {\bf z} c_{ia} + \vec {\bf x} {\Omega}_{ia}{\cdot}
{|{\vec r}|} cos({\Omega}_{ia}t)
+ \vec {\bf y} {\Omega}_{ia}{\cdot}{|{\vec r}|} sin({\Omega}_{ia}t)]}
\end{eqnarray}
At this point a dimensionless parameter $\eta \sim 0.01-0.1$ was 
introduced in order to take into account 
the partial imbalance of electron and ionic currents. 
When $\eta=0$ there exist two co-rotating IAW vortices 
generating the equal currents in a plasma jet 
and this results in a zero net current. 
We assume that $\eta$ is substantially smaller than unity. 
Thus the net current is smaller than steady state 
ion current by order of magnitude or more and it has a 
spatial configuration described by (\ref{current field}).  
Because the distribution functions $W_{i,e}(\vec r, \vec p,t)$ 
\cite{Alexandrov:1984} in the proposed geometry (fig.\ref{fig.1}) 
are expected to be strongly anisotropic and not known yet 
the usage of dynamical equations 
(\ref{pumpwave}-\ref{acouswave1},\ref{contin_eq}) 
with phenomenological parameter $\eta$ 
seems to be reasonable. 
Then substitution of the $nonrelativistic$ 
${{\bf j}_{ia}}$ in a Biot-Savarr integral 
gives the quasistationary magnetic field ${\bf B}$:
\begin{equation}
\label{Biot-Savarr}
{\bf B}(\vec R ,t)=\int {\frac {\mu_0{\:}[{\:}{{\bf j}_{ia}({\vec r},t)} \times (\vec R - \vec r)]}
{4 \pi |\vec R - \vec r|^3}} {\:}{\:}d^{{\:}3}{\vec r}
\end{equation}
The expression for magnetic induction inside the plasma 
solenoid (fig.\ref{fig.1}) and near it's end 
will be published elsewhere. In the far field the time-averaged induction 
$ {\tilde {\bf B}}(\vec R)={{\tau}^{-1} { \int^{\tau}_{0}}{\bf B}(\vec R ,t)}dt$ has the 
configuration of the magnetic dipole, 
whose sign is determined by the optical 
topological charge ${\ell}$:
\begin{equation}
\label{magnetic_dipole}
{\tilde {\bf B}}(\vec R)=
{\frac {\mu_0}{4 \pi}}[{\frac {3 {({\vec \mu}{\:}}{\cdot}\vec R)\vec R}{R^5}}-
{\frac {\vec \mu{\:}}{R^3}}]; {\:} {\:}  {\:} {\:}   \vec \mu \approx 2{\ell} \vec z{\:}
{\frac { {\:} I_{ia} {\:}\pi D^2 L_{jet}}{ \lambda_p}}, {\:}
\end{equation}
where $\vec \mu$ is an effective magnetic moment induced by the 
corkscrew ion-acoustic current $I_{ia}  \approx 
e {\cdot}{\eta \cdot \delta n_{i,e}}{\cdot}{\Omega_{ia}}{\lambda_p}D^2/4 \approx  10^{-3} A{\:}{\:}$ 
for the optical 
fluxes $c \cdot \epsilon_0 |{\bf E}^{o}_{p}|^2 \approx 10^{14} W/cm^2$. The ratio of 
the axial ($|{\tilde {\bf B}}_z| \approx 10^4 {\:}G$) to an  
azimuthal ($|{\tilde{\bf B}}_{\theta}|$) component of a static 
magnetic field is given by: 
\begin{equation}
\label{fields_ratio}
{\frac { |{\tilde{\bf B}}_{z}|}{|{\tilde{\bf B}}_{\theta}|}} \approx
{\frac { e \cdot {\delta n_{e}}|{\bf v}_{\theta}|}{e \cdot {\delta n_{e}}|{\bf v}_{z}|}}= 
{\frac { |{\bf v}_{\theta}|}{|{\bf v}_{z}|}}=
{\frac {{\Omega_{ia}}\cdot R}{v_{ia}}}=
{\frac { {2 \pi D}}{\lambda_p}}
{\:}.
\end{equation} 
This ratio proved to be of kinematic nature and this follows 
from a $solid$ $body$ 
rotation of the charged solenoid (fig.\ref{fig.1}): 
the slow axial current producing 
weak tangential field $|{\tilde{\bf B}}_{\theta}|$ is 
due to IAW translational motion while a fast tangential 
speed $\Omega_{ia} R$ produces 
the large axial induction ($|{\tilde {\bf B}}_z|$).

Apart from the isolated double 
helix interference pattern having diameter near 
$D \approx 100 \mu m$ in the proposed 
loop collider setup (fig.\ref{fig.2}), the more complex experimental situation 
appears in a speckle produced by RPP \cite{Loiseau:2006}. The optical vortices 
in a RPP speckle have a size of $6 \div 100 \mu m$ 
\cite{Tikhonchuk:1997,Loiseau:2006,Basov:1980,
Zeldovich:1985}. Thus the pump field $\bf E_p$ distorted by RPP may produce 
a definite PC component in the Stockes wave $\bf E_s$ reflected from a plasma jet due 
to standard SBS-PC mechanism \cite{Zeldovich:1985}. 
Recently it was pointed out that optical speckle 
consists of randomly spaced set of a 
vortex-antivortex pairs \cite{Okulov:2008J,Okulov:2009}. 
The analogous vortex-antivortex pairs might occur 
in an underdense plasma as well because of a similarity 
of SBS mechanisms in both cases
\cite{Loiseau:2006}. Consequently the rotating charges in a plasma are 
expected to be a sources of the local counter-directed $dc$ magnetic 
dipoles. These dipoles should be 
collocated with a plasma 
vortices having doubled topological 
charge $2 \ell$ \cite{Okulov:2008,Okulov:2009}.
Such essentially $\bf {3D}$ feature of plasma turbulence may be revealed by 
numerical investigation in ($z,x,y,t$) dimension \cite{Najmudin:2001} 
with $\lambda_{ia} \approx \lambda_p/2$ resolution 
instead of low resolution ($z,x,t$) numerical experiments 
in slowly varying envelope approximation \cite{Loiseau:2006}. 
For the sufficiently large values 
of the local magnetic induction (${\bf |{\tilde{B}}|} \sim 10^6 {\:}G$) the 
back action of the magnetic 
field upon plasma currents would be possible. 
Noteworthy also the related task concerning a collision in a plasma 
of the two few cycle femtosecond 
optical vortices with mutually conjugated wavefronts.
\section{Conclusion}
In summary we have shown that the nonrelativistic plasma vortex flow 
resonantly initiated  
by $SBS$ of the two phase-conjugated optical vortices 
is capable to produce a kilogauss quasistatic magnetic field
in a thin slice of the preformed underdense plasma jet \cite{Tikhonchuk:1997}. 
In accordance with our model 
(Eg.\ref{inter_patt1},\ref{density_profile_arb},\ref{current field},\ref{Biot-Savarr}) 
the particles from a preformed 
thermal bath having $KeV$ temperatures 
(e.g. $T_i \approx 0.25{\:}KeV,T_e \approx 0.4 \div 0.7{\:}KeV$ \cite{Tikhonchuk:1997}) 
are accelerated by rotating maxima of light intensity by virtue of ponderomotive force 
and acquire the azimuthal speed 
$ v_{\theta}=\Omega_{ia}R<0.1 \cdot c{\:} $ in addition to axial speed $c_{ia}$ which 
coincides with the IAW velocity. The azimuthal speed 
$ v_{\theta}$ is about two orders of magnitude bigger compared 
to an axial one $v_z=c_{ia}$ (\ref{fields_ratio}).
This provides a corresponding increase of the axial component 
of magnetic induction ${\tilde{\bf B}}_z$ compared 
to the azimuthal one ${\tilde{\bf B}}_{\theta}$. 
The axial tesla-range static magnetic fields proposed here 
in rotating plasma microsolenoid setup might be interesting 
from the point view of studies of the vacuum birefringence phenomena 
where a first results had been obtained quite recently 
in the PVLAS experiment \cite{Shukla:2006} 
(noteworthy the different 
mutual orientation of the laser axis and magnetic induction in our case). 

The case of relativistic intensities and femtosecond laser duration 
${\tau \approx 10^{-14-15}sec} $ in a geometry (fig.\ref{fig.2})
was not studied yet and this case deserve a particular consideration.
Nevertheless it is worth to mention 
the else interesting feature of the 
rotating interference pattern described by (\ref{inter_patt1}). 
For the sufficiently large 
radius $R_{sl}$ the speed of the azimuthal motion $v_{\theta}$ 
of the optical 
interference pattern may reach or even exceed the speed of light 
in a vacuum $\Omega_{ia}R_{sl} \geq c$. 
It well known that such a superluminal motion of the 
interference maxima does not 
violate casuality 
because formula (\ref{inter_patt1}) presumes the 
overlapping of infinitely long pulses.
As is shown previously for the "fast light" phenomena 
\cite{Bolotovskii:2005} the interference maxima can not 
transfer the superluminal signal thus casuality is maintained. 
The same situation holds for the nonlinear laser 
amplifiers \cite{Letokhov:1966, Okulov:1988} and optical pulses 
in dispersive medium \cite{Chiao:1993}. 
In our case when azimuthal speed of rotation 
$v_{\theta}$ approaches $c$, the actual 
relativistic plasma flow will be different from the above formulated 
nonrelativistic SBS model which 
presumes the collocation of the optical intensity maxima and plasma current. 

The partial support of the 
Russian fund for Basic Research Grant 08-02-01229 is acknowledged.

\end{document}